\documentclass[aps,twocolumn]{revtex4}
\usepackage{epsfig,graphics,amssymb,amsmath,subeqnarray}

\def\d{{\rm d}}
\def\p{\partial}
\def\Gm{G_{\rm max}}
\def\Re{{\rm Re}}

\def\k{{\rm Kn}}

\begin{document}

\title{\textbf{A note on the stability of slip channel flows}}
\author{Eric Lauga}
\affiliation{Division of Engineering and Applied Sciences, Harvard University, Cambridge MA 02138, USA}
\author{Carlo Cossu}
\affiliation{LadHyX - Laboratoire d'Hydrodynamique, CNRS-\'Ecole Polytechnique, Palaiseau Cedex 91128, France.}
\date{\today}

\begin{abstract}
We consider the influence of slip boundary conditions on the modal and non-modal stability of pressure-driven channel flows.  In accordance with previous results by Gersting (1974) ({\it Phys. Fluids,} {\bf 17}) but in contradiction with the recent investigation of Chu (2004) ({\it C.R. M\'ecanique}, {\bf 332}), we show that slip increases significantly the value of the critical Reynolds number for linear instability. The non-modal stability analysis however reveals that the slip has a very weak influence on the maximum transient energy growth of perturbations at subcritical Reynolds numbers. Slip boundary conditions are therefore not likely to have a significant effect on the transition to turbulence in channel flows.
\end{abstract}
\maketitle

%...................................................................
%CC Below I would cut most of the references
The advances in microfabrication techniques using polymeric or silicon-based materials has allowed  to gain significant understanding on the behavior of fluids at small scales  \cite{ho98,stonereview,squiresquake}. One topic of current interest concerns the validity of the no-slip boundary condition for Newtonian liquids near solid surfaces \cite{goldstein,vinogradova99,granick03,tabeling04,laugareview}. A large number of recent experiments on small scales with flow driven by pressure gradients, drainage, shear,  or electric field  have reported an apparent breakdown of the no-slip condition, with slip lengths possibly as large as microns. The slip length, $\lambda$, is defined as the ratio of the surface velocity to the surface shear rate; $\lambda=0$ corresponds to a no-slip condition, and $\lambda=\infty$ to a perfectly slipping surface.

%\cite{schnell56,churaev84,churaev99,watanabe99,cheng02,meinhart02,breuer03,lumma03}\cite{baudry01,cottin02,granick01,craig01,neto03,henry04,bonaccurso02,bonaccurso03,sun02,cho04},  \cite{pit00} \cite{churaev02}
 
Since the transition to turbulence in wall-bounded flows occurs at large values of the Reynolds number, studies in shear-flow instabilities have usually been outside the realm of microfluidics. However, a set of recent investigations of the linear modal stability of pressure-driven flows in two-dimensional channels \cite{chu00,chu03,chu04} has reported that slip boundary conditions decrease the critical Reynolds number, from $\Re=5772$ (its classical no-slip value obtained for Poiseuille flow) to $\Re\approx 100$, in strong disagreement with early calculations of Gersting \cite{gersting74}.  Such results would potentially  have a major impact on both turbulence and microfluidic studies. 

The goal of this note is twofold. First, we resolve the disagreement between the above cited results. A careful analysis of the derivation of the equations used in Refs.~\cite{chu00,chu03,chu04} reveals that incorrect slip boundary conditions on the perturbations were used in the modal stability analysis. The use of the correct appropriate boundary conditions on the perturbations reveals the strongly stabilizing effect of slip on the eigenvalues of the linear stability operator, confirming earlier results  \cite{gersting74}. Recent advances in the domain of shear flow instabilities have however revealed the usual lack of relevance of modal stability analysis, contrasted to non-modal stability analysis, in subcritical transition in channel flows (for a review see, e.g.,  Refs.~\cite{trefethen93,schmid01}). The second goal of this note is therefore to quantify the effect of slip on the non-modal stability of viscous channel flows. To this end, we compute the maximum transient energy growth \cite{butler92} in the presence of slip at subcritical Reynolds numbers. We find that, for all the considered combinations of streamwise and spanwise wavenumbers, the  effect of slip on the maximum energy growth and on the associated optimal perturbations is  weak.

%...................................................................
\paragraph*{Problem setting.}

We consider the flow  between two parallel plates located at $y^*=\pm h$ of a fluid with shear viscosity $\mu$ driven by a constant pressure gradient $\d p^* / \d x^*$ in the $x^*$-direction. If  we non-dimensionalize lengths by $h$,  velocities by $U_{\rm ref} = h^2(-\d p^*/\d x^*)/2\mu$, time by $h/U_{\rm ref}$ and pressure by $\rho U_{\rm ref}^2$, the dimensionless incompressible Navier-Stokes equations for the velocity and pressure fields, $({\bf u},p)$,  read
\begin{equation}\label{NS}
\left(\frac{\p }{\p t} + {\bf u}\cdot \nabla \right){\bf u} =  -\nabla p +\frac{1}{\Re}\nabla^2 {\bf u} ,\quad \nabla \cdot {\bf u}=0,
\end{equation}
where we have defined  the Reynolds number for this flow as $\Re={\rho h U_{\rm ref}}/{\mu}$. 
We assume in this paper that the flow satisfies  slip boundary conditions on both surfaces, with slip lengths $\lambda_1$ and $\lambda_2$ at $y=h$ and $y=-h$, respectively. If we define the Knudsen numbers $\k_1=\lambda_1/h$ and $\k_2=\lambda_2/h$, and denote by ($u,v,w$)  the streamwise, wall normal and spanwise components of ${\bf u}$, the boundary conditions for Eq.~\eqref{NS} are $v=0$ at $y=\pm 1$  and
\begin{subeqnarray}\label{slipbc}
\displaystyle u+\k_1\frac{\p  u}{ \p y} &=& \displaystyle w+\k_1\frac{\p  w}{ \p y} =0, \quad  y=1, \\
\displaystyle \quad u-\k_2\frac{\p u}{ \p y}  &=& \displaystyle   w-\k_2\frac{\p w}{ \p y} =0, \quad  y=-1.
\end{subeqnarray}

%...................................................................
\paragraph*{Linear stability.}

We are interested  in the stability of the steady unidirectional base flow $ {\bf U}= U(y){\bf e}_x$ satisfying Eqs.~\eqref{NS} and \eqref{slipbc}, 
\begin{eqnarray}\label{base}
 U({ y})&=&1+\frac{2(\k_1+\k_2+2\k_1\k_2)}{2+\k_1+\k_2}  \\
&+& \left(\frac{2(\k_1-\k_2)}{2+\k_1+\k_2} \right)  y- y^2.
\end{eqnarray}
In the absence of slip, $\k_1=\k_2=0$, and Eq.~\eqref{base} reduces to the standard Poiseuille solution $ U(y)=1-y^2$.
In order to characterize the stability of Eq.~\eqref{base}, we  write the total velocity field as the sum of the base flow plus small perturbations, ${\bf u}= {\bf U} + {\bf u}'$, $p= P+p'$, and linearize the Navier-Stokes equations around $({\bf U}, P)$. 
This procedure is classic and we refer, e.g., to Refs.~\cite{drazin81,schmid01} for the details.
The same standard procedure is applied to the boundary conditions (Eq.~\ref{slipbc}). These linear boundary conditions are satisfied by both the total flow $\{u= U+u',v=v',w=w'\}$ and the base flow itself $\{ U,0,0\}$. Consequently, a simple subtraction shows that the boundary conditions for the perturbations are also of the form of Eq.~\ref{slipbc}. These boundary conditions are the same as those used by Gersting in his stability analysis \cite{gersting74} and differ from the incorrect boundary conditions used in Refs.~\cite{chu00,chu03,chu04} that implicitly assume ${\bf u}'={\bf 0}$ at $y=\pm 1$. Therefore, in Refs.~\cite{chu00,chu03,chu04},  slip boundary conditions are assumed for the basic flow but no-slip boundary conditions are used for the perturbations, leading to incorrect results.

\begin{figure}[t]
\centering
\includegraphics[width=.35\textwidth]{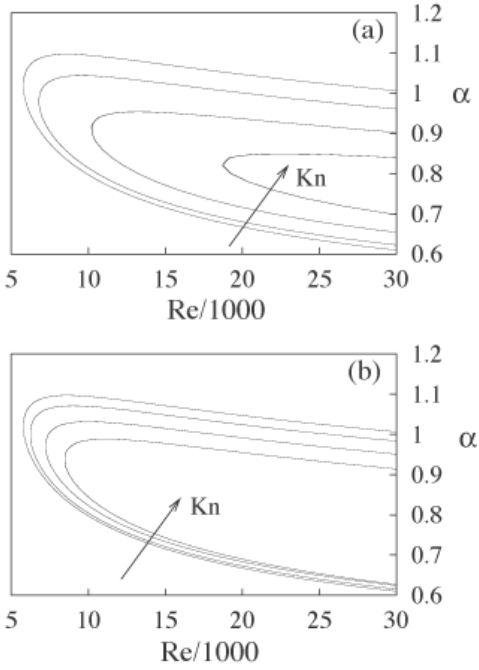}
\caption{Neutral curve $\omega_i(\alpha,\beta,\Re)=0$ for the symmetric slip case (a) and asymmetric slip case (b), $\beta=0$, and values $\k=$ 0 (no-slip), 0.01, 0.02, and 0.03.} 
\label{NeutCurv}
\end{figure}

Following a standard procedure (see, e.g., Ref.~\cite{schmid01} for details), the linearized Navier-Stokes equations are recast in a set of two differential equations for the wall-normal velocity, $v'$, and the wall-normal vorticity, $\eta'= \p u'/\p z - \p w'/\p x$.  Exploiting the homogeneous nature of the streamwise and spanwise directions, perturbations are Fourier-transformed in the form
\begin{equation}
{\bf u'}(x,y,z,t)=\widehat {\bf u}(\alpha,y,\beta,t)\,{\rm e}^{i(\alpha x +\beta z)},
\end{equation}
and therefore, $\eta'(x,y,z,t)=\widehat \eta (\alpha,y,\beta,t)\,{\rm e}^{i(\alpha x +\beta z)}$,
with $\widehat\eta  =i\beta \widehat u - i\alpha \widehat w$.
The standard evolution equation for $(\widehat v,\widehat \eta )$, is  finally obtained to be \cite{schmid01}
\begin{equation}
\frac{\p }{\p t}\left(
\begin{array}{c}
\Delta \widehat v \\
\widehat \eta
\end{array}\right)
 = 
 \left(
\begin{array}{cc}
{\cal L}  & 0\\
{\cal C} & {\cal S}
\end{array}\right)
\cdot
 \left(
\begin{array}{c}
\widehat v \\
\widehat \eta
\end{array}\right),
\label{OSS}
\end{equation}
where the operators are defined as
\begin{eqnarray}
{\cal L} & \triangleq  & -i\alpha  U \Delta + i \alpha D^2   U + \Delta(\Delta / \Re),\\
{\cal C} & \triangleq  & -i\beta D   U,\\
{\cal S} & \triangleq  & -i \alpha  U + \Delta / \Re,
\end{eqnarray}
with $\Delta\triangleq D^2-\alpha^2-\beta^2$, where $D$ denotes derivatives with respect to $y$
The fourth-order system of equations, Eq.~\eqref{OSS}, requires boundary conditions for both $\widehat v$ and $\widehat \eta$. Using the continuity equation, $i \, \alpha \widehat{u} +  D \widehat{v}+ i\, \beta \widehat{w}=0$ together with the boundary conditions in Eq.~\eqref{slipbc},  it is straightforward to show that the boundary conditions for $(\widehat{v},\widehat{\eta})$ are
\begin{subeqnarray} \label{OSSBC}
\label{bcv_1} \widehat{v}&=&D \widehat{v}+\k_1 D^2\widehat{v}=0, \quad y=1,\\
\label{bcv_2}  \widehat{v}&=&D\widehat{v}-\k_2 D^2\widehat{v}=0, \quad y=-1,\\
\label{bcomega_1} && \widehat{\eta}+\k_1 D\widehat{\eta}=0, \quad y=1,\\
\label{bcomega_2} &&\widehat{\eta}-\k_2  D\widehat{\eta}=0, \quad y=-1,
\end{subeqnarray}
and for simplicity, we restrict the analysis in this note to symmetric slip ($\k_1=\k_2=\k$) and asymmetric slip cases ($\k_1=\k$, $\k_2=0$). We emphasize again that these boundary conditions are different from those used in Refs.  \cite{chu00,chu03,chu04}, where instead $\k_1$ and $\k_2$ were set to zero in Eq.~\eqref{bcv_1}.

\begin{figure}[t]
\centering
\includegraphics[width=.35\textwidth]{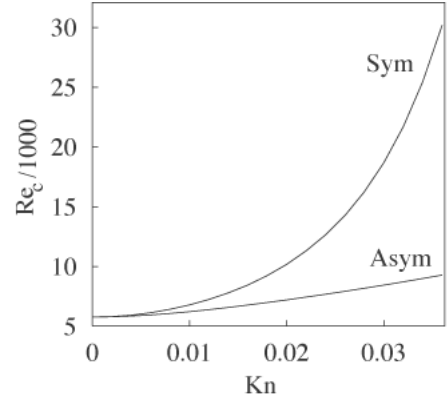}
\caption{Critical Reynolds number for linear stability $\Re(\k)$ for the symmetric   and asymmetric slip cases.}
\label{Recrit}
\end{figure}

%...................................................................
\paragraph*{Numerical method.}
A Chebyshev collocation method is used to discretize the system, Eq.~\ref{OSS}, and standard methods (described in Ref.~\cite{schmid01} and references therein) are then employed to compute eingenvalues, eigenmodes and transient energy growth. The standard implementation of these methods is modified by changing the standard homogeneous no-slip boundary conditions into the more general slip boundary conditions (Eq.~\ref{bcv_1}). All the results presented below have been obtained with $97$ collocation points. Convergence of the results had been verified and the code has been thoroughly tested by comparing both the modal and the non-modal results in the case of no-slip  \cite{schmid01}, as well as with the modal symmetric slip results reported in Ref.~\cite{gersting74}.

%...................................................................

\paragraph*{Influence of slip on modal stability.}
The modal stability analysis assumes solutions in the form of normal modes, 
$\{\widehat   v, \widehat   \eta\} (\alpha,y,\beta,t)=
 \{\widetilde v, \widetilde \eta\} (\alpha,y,\beta,\omega)\,{\rm e}^{-i \omega t}$, where the complex frequency, $\omega$, is the solution to an eigenvalue problem, 
which is solved numerically.  The flow is found to be linearly unstable if there exists at least one eigenvalue with positive imaginary part, $\omega_i >0$. The Squire theorem \cite{drazin81} applies to this flow  and the critical modes are two-dimensional ({\it i.e.}, with $\beta=0$). The neutral curve $\omega_i(\alpha,\beta=0,\Re)=0$ in the symmetric slip case is displayed in Fig.~\ref{NeutCurv}a. Boundary slip is found to significantly shift the neutral curve towards larger values of the Reynolds number, indicating a strongly stabilizing influence of slip on linear stability.
%(note that we restrict the analysis to small values of $\k$ to be consistent with  the large values of  $\Re$). 
Results for the asymmetric slip case, displayed in Fig.~\ref{NeutCurv}b, are similar, although less pronounced. The dependence of the critical Reynolds number for linear stability,  $\Re_c$, with the Knudsen number, $\k$, is diplayed in Fig.~\ref{Recrit} and confirms the stabilizing effect of slip on shear-flow instabilities. Our results, which use the correct boundary conditions, Eq.~\eqref{bcv_1}, agree with the symmetric slip calculations of Ref.~\cite{gersting74}, but, as expected, are in strong contradiction with the conclusions reported in Ref.~\cite{chu00,chu03,chu04}.

\begin{figure}[t]
\centering
\includegraphics[width=.35\textwidth]{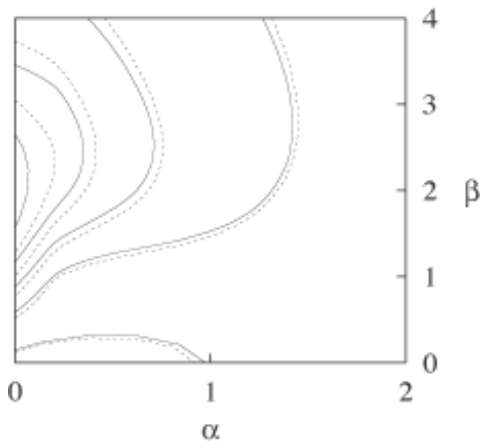}
\caption{Map of the iso-values of the transient energy growth $\Gm(\alpha,\beta)$ for $\Re=1500$ in two cases: No-slip (solid line) and symmetric slip boundary conditions with $\k=0.03$ (dashed line). The values of $\Gm$ are 10, 100, 200, 300 and 400 from the outer to the inner curve.}
\label{TransientMap}
\end{figure}

%...................................................................
\paragraph*{Non-modal stability analysis.}
In the absence of slip at the walls, the Poiseuille flow is known to undergo transition to turbulence at Reynolds numbers well below the critical Reynolds number corresponding to the onset of linear modal instability. This strongly subcritical transition scenario has been related to the strongly non-normal nature of the linearized operator (Eq.~\ref{OSS}), explaining the potential of the flow to sustain large transient energy growth, possibly triggering the transition to turbulence for values of the Reynolds number much smaller than $\Re_c$ \cite{trefethen93,schmid01}.  The standard modal stability analysis is therefore extended to the non-modal (or generalized \cite{farrell96}) stability analysis where, for instance, the maximum transient energy growth is computed.  
Let us define, for a given Fourier mode, the instantaneous kinetic energy of the flow perturbations as
\begin{equation}
E(t,\alpha,\beta,\widehat{\bf u}_0)\triangleq \int_{-1}^{1} |\widehat{\bf u}(\alpha,y,\beta,t)|^2\d y,
\end{equation}
which is a function of time and the initial condition, $\widehat{\bf u}_0 \triangleq \widehat{\bf u}(\alpha,y,\beta,0)$. If we denote by $G(t)$ the energy growth at time $t$, maximized over all non-zero initial conditions, 
\begin{equation}
G(t,\alpha,\beta)=\max_{\widehat{\bf u}_0\neq 0} \left(\frac{E(t,\alpha,\beta,\widehat{\bf u}_0)}{E(0,\alpha,\beta,\widehat{\bf u}_0)}\right),
\end{equation}
then the maximum transient energy growth possible over all times, $\Gm(\alpha,\beta)$, is defined as
\begin{equation}
\Gm(\alpha,\beta)=\max_{t\geq 0} G(t,\alpha,\beta).
\end{equation}

\begin{figure}[t]
\centering
\includegraphics[width=.35\textwidth]{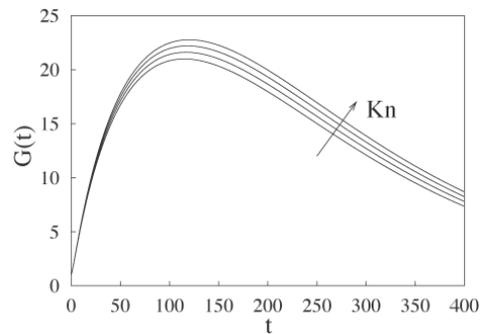}
\caption{Evolution in time of the initial condition leading to the largest finite-time energy amplification $\Gm$  for the Fourier mode $(\alpha,\beta)=(0,2)$, $\Re=1500$, and in the symmetric slip case  with $\k=$0 (no-slip), 0.01, 0.02 and 0.03.}
\label{TransientRe1500}
\end{figure}

In Fig.~\ref{TransientMap} we report the iso-values of  $\Gm(\alpha,\beta)$ computed for $\Re=1500$  for both the no-slip (solid line) and the symmetric slip case (dashed line). Although the maximum energy growth with slip is always larger than in the case of no-slip, the increase is small and therefore slip hardly affects transient energy growth. The maximum energy growth is obtained for $\alpha=0$ and $\beta=2$ for both slip and no-slip boundary conditions.   Fig.~\ref{TransientRe1500}  displays the time evolution of the optimal energy growth  $G(t,\alpha=0,\beta=2)$, at $\Re=1500$ and in the symmetric slip case, for  different values of the Knudsen number. The small increase of the optimal growth with $\k$ appears in all cases. Furthermore, the time where the maximum growth is attained is also slightly increased by the slip.  As both the square root of maximum growth and the time at which it is attained depend linearly on the Reynolds numbers, these effects suggest that the effect of slip induces an increase of an effective Reynolds number, which is consistent with the observation that slip flows have, for the same forcing, a larger flow rate than in the case of no-slip.

\begin{figure}[t]
\centering
\includegraphics[width=.35\textwidth]{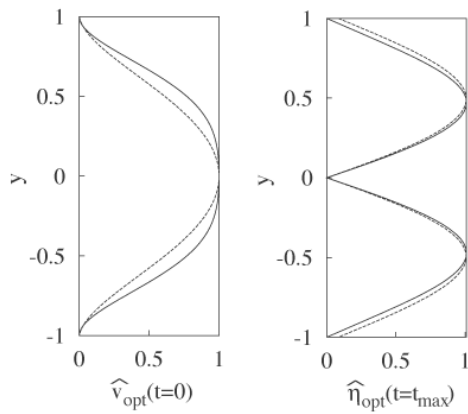}
\caption{Optimal initial condition, $\widehat{v}_{\rm opt}(t=0)$, and optimal response, $\widehat{\eta}_{\rm opt}(t=t_{\rm max})$, leading to largest transient energy growth at $Re=1500$ and for $(\alpha,\beta)=(0,2)$, for both no-slip (solid line) and symmmetric slip with $\k=0.03$ (dashed line) boundary conditions.}
\label{worst}
\end{figure}

In the case of no-slip channel flow it is known that the initial perturbations inducing the largest energy growth are streamise vortices, while the most amplified response consist in streamwise streaks. Translated in terms of the $v'$ and $\eta'$ variables, this means that the optimal initial perturbations are of $v$-type, with $\eta$ negligible, while, on the contrary, the most amplified response is of $\eta$-type, with $v$ negligible. This is also the case with slip boundary conditions. In Fig.~\ref{worst} we reproduce, for $\Re=1500$, the optimal initial condition, $\widehat v_{\rm opt}(y,t=0)$ (left), and the optimal response $\widehat \eta_{\rm opt}(y,t=t_{\rm max})$ (right), corresponding to the largest transient energy growth, $\Gm(\alpha=0,\beta=2)$. The shape of the optimal initial perturbation differ slightly from the no-slip case, while the optimal responses are nearly undistiguishable, except near the wall, where the effect of the slip boundary conditions is apparent. The lift-up mechanism, by which low amplitude vortices are converted into large amplitude streaks seems therefore to be only slightly sensitive to slip boundary conditions at the wall. Similar results are obtained for other values of ($\k$, $\Re$) and for asymmetric slip boundary conditions.

\section*{Acknowledgements}
Funding by the Office of Naval Research and the Harvard MRSEC is gratefully acknowledged (EL).

\bibliographystyle{unsrt}
\bibliography{bib_thesis_elauga}
\end{document}